\documentclass[aps,prd,unsortedaddress,
  showpacs,preprintnumbers]{revtex4-1} 

\usepackage{graphicx}
\usepackage{dcolumn}
\usepackage[dvips]{color}
\newcommand{\be}{\begin{equation}}
\newcommand{\ee}{\end{equation}}
\newcommand{\bea}{\begin{eqnarray}}
\newcommand{\eea}{\end{eqnarray}}

\newcommand{\bfk}{\mbox{\boldmath $k$}}
\newcommand{\bfq}{\mbox{\boldmath $q$}}

\newcommand{\kt}{k_\perp}

\def\lsim{\mathrel{\rlap{\lower4pt\hbox{\hskip1pt$\sim$}}\raise1pt\hbox{$<$}}}
\def\gsim{\mathrel{\rlap{\lower4pt\hbox{\hskip1pt$\sim$}}\raise1pt\hbox{$>$}}}

\begin{document}


\title{Transverse single spin asymmetry in 
{\boldmath$e+p^\uparrow \to  e+J/\psi +X $ and transverse momentum dependent 
evolution of the Sivers function\\}}

\author{Rohini M. Godbole}
\affiliation{Centre for High Energy Physics, Indian Institute of Science, Bangalore, India.}
\email{rohini@cts.iisc.ernet.in}

\author{Anuradha Misra}
\affiliation{Department of Physics, University of Mumbai, Mumbai, India.}
\email{misra@physics.mu.ac.in}

\author{Asmita Mukherjee}
\affiliation{Department of Physics, Indian Institute of Technology Bombay, Mumbai, India.} 
\email{asmita@phy.iitb.ac.in}

\author{Vaibhav S. Rawoot}
\affiliation{Department of Physics, University of Mumbai, Mumbai, India.}
\email{vaibhavrawoot@gmail.com}

\date{\today}
\begin{abstract}

We extend our analysis  of transverse single spin asymmetry (SSA) in 
electroproduction of $J/\psi $ to include the effect of
the scale evolution of the transverse momentum dependent 
(TMD) parton distribution functions and gluon Sivers function. 
We estimate SSA for JLab, HERMES, COMPASS, and eRHIC energies using the
color evaporation model of charmonium production, using an analytically obtained 
approximate solution of  TMD evolution equations discussed in the literature. We find that there is a reduction in the asymmetry 
compared with our predictions for the earlier case considered by us,
wherein the $Q^2$ dependence came only from  DGLAP evolution of the 
unpolarized gluon densities and a different parametrization
of the TMD Sivers function was used. 

\end{abstract}

\maketitle

\section{\label{sec:level1}Introduction}

The factorization theorems of quantum chromodynamics (QCD), which enable us to calculate 
the cross sections of hadronic processes as a convolution of partonic cross section with parton 
distribution functions(PDFs)  and fragmentation functions(FFs), are based on a collinear approximation. In this 
approximation, the intrinsic transverse momentum is integrated over in the definition of PDFs and FFs. 
However, this collinear factorization at leading twist is unable to account for the single spin asymmetries (SSAs) 
that have been observed in processes involving scattering off polarized hadrons \cite{AdamsBravar1991,KruegerAllogower1999,Hermes,Compass, alesio-review}. 
One of the approaches that has been used to explain these SSAs is based on the 
transverse momentum dependent TMD factorization formalism wherein the PDFs and FFs depend on intrinsic 
transverse momentum ${\bfk_\perp} $ also in addition to the momentum fraction variable $x$ \cite{tmd-fact1, Sivers1990, tmd-fact2}. 
TMD factorization is particularly useful for describing the processes that are sensitive to the parton's 
intrinsic transverse momentum. A formalism for TMD factorization has been developed by Collins and Soper \cite{fact} 
and has been used to study Drell-Yan (DY), semi-inclusive deep inelastic scattering (SIDIS) 
and back-to-back  hadron production in $e^+ e^-$ annihilation at small transverse momentum \cite{Ji:2005}. 
There has been huge interest amongst both theoreticians and experimentalists
in understanding to what extent the above factorization holds, and in the study  of the TMD PDFs, 
TMD FFs, and the resulting SSA's with the aim of 
probing  the transverse momentum distribution of partons inside hadrons and the spin structure of hadrons.  
Henceforth, we will refer to TMD PDFs and TMD FFs collectively as ``TMDs.``

One of the important TMDs is the Sivers function which describes the transverse momentum distribution of an 
unpolarized parton inside a transversely polarized hadron. Estimates of Sivers asymmetry have been given in the
Drell Yan (DY), SIDIS and D-meson production based on a 
generalized factorization formula involving TMD PDFs and TMD FFs. In a previous work \cite{Godbole:2012bx}, we proposed $J/\psi$ 
production as a probe of gluon Sivers function and gave estimates for  SSAs in low virtuality electroproduction of $J/\psi$ 
at JLab, HERMES, COMPASS, and eRHIC energies. 
The use of heavy quark and quarkonium production to get information about the gluon densities, as well as 
about the underlying QCD dynamics involved in the formation of the $Q \bar Q$ bound states, 
has been explored for a variety of beams, polarized and unpolarized, as well as a variety of targets~\cite{oldrefs}. 
At leading order (LO), the charmonium production receives a contribution only from a single partonic 
subprocess $\gamma g \rightarrow c {\bar c}$ . Hence,  SSA in 
$e + p^\uparrow \rightarrow e +J/\psi +X$ can be used as a clean 
probe of gluon Sivers function.
There exist three different models for charmonium production.
In the color singlet model \cite{sing} 
the cross section for charmonium production is factorized into a short distance part for 
$c {\bar c}$ pair production 
calculable in perturbation theory and a nonperturbative matrix element
for the formation of a bound state, which is produced in a color singlet state.
In the color evaporation model, first proposed by Halzen and Matsuda 
\cite{hal} and Fritsch \cite{fri}, a statistical treatment of color is made and 
the probability of finding a specific quarkonium state is assumed to be independent 
of the color of the heavy quark pair. In later versions of this model, it has been found
that the data are better fitted if a phenomenological factor is included in the differential 
cross section formula, which depends on a Gaussian distribution of the transverse 
momentum of the charmonium \cite{ce2}.
A more recent model of charmonium production is the color octet model \cite{octet}.
This is based on a factorization approach in nonrelativistic QCD (NRQCD), and it allows $c {\bar c}$ pairs   
to be produced in color octet states. Here again, one 
requires knowledge of the nonperturbative color octet matrix elements,
which are determined  through fits to the data on charmonium production. 
The investigation of SSAs in charmonium production as a 
possible  tool towards resolution of the charmonium production mechanism puzzle 
has been discussed within the NRQCD framework some time back \cite{feng}. 
It was shown in Ref.~\cite{feng}, using general arguments, that SSA in heavy quarkonium 
production is sensitive to the color configuration of the $ q{\bar q} $ pair produced at a short distance. Based on an analysis 
of initial and final state interactions , the authors argued that in the
NRQCD framework, asymmetry is nonzero in $e$-$p$ collisions  only in color 
octet model and in $p$-$p$ collisions it is non-zero only in the color singlet model.

In our earlier work in Ref. \cite{Godbole:2012bx} we had performed a   
phenomenological study of the SSA in charmonium production.
We had calculated  the asymmetry  in the process $e+p^\uparrow \to  e+J/\psi +X $ 
using the color evaporation model of charmonium production \cite{cem0}.
As discussed therein, in this exploratory study of SSA as a probe of the gluon Sivers function, 
we had chosen  to work with CEM due to its simplicity.   In the older work,  we had used parameters of Sivers function 
fitted to SIDIS data in  Ref.~\cite{Anselmino:2011gs}. Earlier 
parametrizations of TMDs like the  ones we had used
\cite{Godbole:2012bx} had been obtained by assuming a 
parton model picture of TMD factorization and directly 
fitting~\cite{Anselmino:2011gs,Anselmino:2008sga} the calculated cross sections to 
experimental data from HERMES and COMPASS.  However, the $Q^2$ - evolution of TMDs had 
not been taken into  account in obtaining these parametrizations. 
Our earlier  work that used these thus could not include the effect of 
evolution of TMDs. 

In the present work, we improve our previous estimates by taking into account 
the $Q^2$ evolution of TMD PDFs and the Sivers function, and energy dependence of the Sivers asymmetry has attracted a lot of
attention in the recent past \cite{Collins:2011book, Aybat:2011zv, Aybat:2011ge, Aybat:2011ta, Boer:2013zca, Sun:2013dya}.
A TMD factorization framework taking into account the $Q^2$ evolution of 
TMDs has been proposed \cite{Collins:2011book}. This framework has been 
used to obtain evolved TMDs from fixed scale fits. 
A  simple  strategy for extracting the TMD evolved Sivers function has been proposed, and a best fit of 
the SIDIS asymmetries has been performed taking into account the 
$Q^2$ dependence of TMDs~ \cite{Anselmino:2012aa} using data 
from Ref.~\cite{:2009ti}. The authors have also compared these fits with 
those  obtained earlier~\cite{Anselmino:2011gs}. In fact, the TMDs evolve 
much faster than the integrated PDFs, for which the scale evolution is 
given by the DGLAP equation.  In this work, we apply the formalism of 
Ref. \cite{Anselmino:2012aa} to $J/\psi$ electroproduction to improve our 
earlier estimates, taking into account the effect of $Q^2$ evolved TMDs.
In the current investigation, we restrict ourselves to the use of analytical
formulas for the evolved TMDs, given in Ref.~\cite{Anselmino:2012aa}.

In Sec. II, we review the formalism for $J/\psi$ production in CEM and 
also summarize the TMD evolution formalism of  Ref.~\cite{Anselmino:2012aa}. 
In Sec. III, we present our estimates for asymmetry obtained by taking 
into account the TMD evolution and  scale dependence of the Gaussian width of 
PDFs. Section IV contains the summary and discussion of our results. 

\section{FORMALISM }

\subsection{Asymmetry in $J/\psi$ production}

We consider the LO parton model cross section for low virtuality electroproduction (photoproduction)
of $J/\psi$ within the color evaporation model. According to CEM, the cross section for charmonium production is proportional to the
rate of production of the $c\bar{c}$ pair integrated over the mass range $2m_c$ to $2m_D$ \cite{cem0}
\be
\sigma=\frac{1}{9}\int_{2m_c}^{2m_D} dM_{c\bar{c}} \frac{d \hat \sigma_{c\bar{c}}}{dM_{c\bar{c}}}
\label{sigmacem}
\ee  
where $m_c$ is the charm quark mass and $2m_D$ is the $D\bar{D}$ threshold.
The partonic cross section $\displaystyle\frac{d\hat{\sigma}_{c\bar{c}}}{dM_{c\bar{c}}}$ is calculable perturbatively, 
$M_{c\bar{c}}$ being the invariant mass of the $c\bar{c}$ pair.
The differential cross section for $\gamma+p\rightarrow J/\psi+X$ is given by
\be
\frac{d\sigma^{\gamma p\rightarrow c\bar{c}}}{dM^2_{c\bar{c}}}=
\int d{x_g} f_{g/p}(x_g) \frac{d\hat{\sigma}^{\gamma g\rightarrow c\bar{c}}}{dM^2_{c\bar{c}}}
\label{xsec-ccbar}
\ee
where $f_{g/p}(x)$ is the gluon distribution in the proton. 

Using the Weizsacker-Williams approximation \cite{wwf}, one can convolute the 
cross section given by Eq. (\ref{xsec-ccbar}) with a photon flux factor to obtain the electroproduction cross section 
for $e+p\rightarrow e+ J/\psi+X$ 
\be
\frac{d\sigma^{ep\rightarrow e+ J/\psi+X}}{d M_{c\bar{c}}^2}=\int d{x_\gamma}\> f_{\gamma/e}(x_\gamma)\> 
\frac{d\hat{\sigma}^{\gamma p\rightarrow c\bar{c}}}{d M_{c\bar{c}}^2}
\label{xsec-ep}.
\ee

In the above, $f_{\gamma/e}(x_\gamma)$ is the distribution function of the photon in 
an electron with  $x_\gamma$ denoting the energy fraction of the electron that the photon carries. 
We use the expression for $f_{\gamma/e}(x_{\gamma})$ from Refs.~\cite{ww1,ww}   given by
\bea
f_{\gamma/e}(x_\gamma,E)=\frac{\alpha}{\pi} \{\frac{1+(1-x_\gamma)^2}{x_\gamma}\left(ln\frac{E}{m}-\frac{1}{2}\right)
+\frac{x_\gamma}{2}\left[ln\left(\frac{2}{x_\gamma}-2\right)+1\right] \nonumber \\
+\frac{(2-x_\gamma)^2}{2x_\gamma}ln\left(\frac{2-2x_\gamma}{2-x_\gamma}\right) \},
\label{ww-knihel}
\eea
where $x_\gamma$ is the energy fraction of the electron carried by the photon
, $m$ is the mass of the electron, and $E$ is the energy of the electron.

Thus, the cross section for electroproduction of $J/\psi$ using  WW approximation is given by
\be
\sigma^{ep\rightarrow e+J/\psi+X}=
\int_{4m_c^2}^{4m_D^2} dM^2_{c\bar{c}} \int d{x_\gamma}\> dx_{g}\> f_{\gamma/e} (x_\gamma)\> f_{g/p}(x_g) 
\>\frac{d\hat{\sigma}^{\gamma g\rightarrow c\bar{c}}}{dM_{c\bar{c}}^2}.
\label{xsec-gammap}
\ee

To calculate SSA in the scattering of electrons off a polarized proton target, we assume
a generalization of the CEM expression by taking into account the transverse momentum dependence
of the Weizsacker-Williams function and gluon distribution function \cite{Godbole:2012bx}
\be
\frac{d\sigma^{e+p^\uparrow\rightarrow e+J/\psi + X}}{d M^2}=
\int  dx_\gamma\> dx_g\> [d^2\bfk_{\perp\gamma}d^2\bfk_{\perp g}]\>
f_{g/p^{\uparrow}}(x_{g},\bfk_{\perp g})
f_{\gamma/e}(x_{\gamma},\bfk_{\perp\gamma})
\frac{d\hat{\sigma}^{\gamma g\rightarrow c\bar{c}}}{dM^2}
\label{dxec-ep}
\ee
where $M^2\equiv M_{c\bar{c}}^2$. 
The difference in $d\sigma^\uparrow$ and $d\sigma^\downarrow$ is parametrized in terms of the gluon
Sivers function
\be
d\sigma^\uparrow-d\sigma^\downarrow=
\int dx_\gamma\> dx_g\> d^2\bfk_{\perp\gamma}\>d^2\bfk_{\perp g}\>
\Delta^{N}f_{g/p^{\uparrow}}(x_{g},\bfk_{\perp g})\>
f_{\gamma/e}(x_{\gamma},\bfk_{\perp\gamma})\>
{d\hat{\sigma}}^{\gamma g\rightarrow c\bar{c}}
\label{dxsec-nssa}
\ee
where $d\hat{\sigma}$ is the elementary cross section for the process $\gamma g\rightarrow c\bar{c}$
given by

\be
d\hat\sigma = \frac{1}{2\hat s} \> \frac{d^3p_c}{2E_c} \frac{d^3p_{\bar{c}}}{2E_{\bar{c}}}\>  
\frac{1}{(2\pi)^2} \> \delta^4(p_\gamma + p_g - p_c - p_{\bar{c}}) \>
\overline{\left\vert \, M_{\gamma g \to c\bar{c}} \, \right\vert^2} \>. 
\label{ecs}
\ee

Following the procedure in Ref.\cite{Godbole:2012bx}, we  obtain 
\bea
\frac{d^{4}\sigma^\uparrow}{dydM^2d^2\bfq_T}-\frac{d^4\sigma^\downarrow}{dydM^2d^2\bfq_T}=
\frac{1}{s}\int [d^2\bfk_{\perp\gamma}d^2\bfk_{\perp g}]
\Delta^{N}f_{g/p^{\uparrow}}(x_{g},\bfk_{\perp g})
f_{\gamma/e}(x_{\gamma},\bfk_{\perp\gamma}) \nonumber\\
\times\>\delta^2(\bfk_{\perp\gamma}+\bfk_{\perp g}-\bfq_T)
\hat\sigma_{0}^{\gamma g\rightarrow c\bar{c}}(M^2)
\label{num-ssa}
\eea

and 

\bea
\frac{d^{4}\sigma^\uparrow}{dydM^2d^2\bfq_T}+\frac{d^4\sigma^\downarrow}{dydM^2d^2\bfq_T}=
\frac{2}{s}\int [d^2\bfk_{\perp\gamma}d^{2}\bfk_{\perp g}]
f_{g/p}(x_g,\bfk_{\perp g})
f_{\gamma/e}(x_{\gamma},\bfk_{\perp\gamma}) \nonumber\\
\times\>\delta^2(\bfk_{\perp\gamma}+\bfk_{\perp g}-\bfq_T)
\hat\sigma_{0}^{\gamma g\rightarrow c\bar{c}}(M^2)
\label{den-ssa}
\eea

where 
\be
x_{g,\gamma} = \frac{M}{\sqrt s} \, e^{\pm y}
\label{x-gammag}
\ee

and the partonic cross section is given by \cite{gr78}
\be
\hat{\sigma_0}^{\gamma g\rightarrow c\bar{c}}(M^2)=
\frac{1}{2}e_{c}^2\frac{4\pi\alpha\alpha_s}{M^2}
[(1+ v -\frac{1}{2} v^2)\ln{\frac{1+\sqrt{1- v}}{1-\sqrt{1-v}}}
-(1+v)\sqrt{1-v}].
\label{jpsics}
\ee
Here, $\displaystyle v=\frac{4 m_c^2}{M^2}$ and $M^2\equiv\hat{s}$.

Integrating Eqs. (\ref{num-ssa})  and (\ref{den-ssa}) over $M^2$, 
we obtain  the difference and sum of $\displaystyle\frac{d^{3}\sigma^\uparrow}{dyd^2\bfq_T}$ 
and $\displaystyle\frac{d^3\sigma^\downarrow}{dyd^2\bfq_T} $for $J/\psi$ production. 

The Sivers asymmetry is defined as 
\be
A_N^{\sin({\phi}_{q_T}-\phi_S)} =\frac{\int d\phi_{q_T}
[d\sigma ^\uparrow \, - \, d\sigma ^\downarrow]\sin({\phi}_{q_T}-\phi_S)}
{\int d{\phi}_{q_T}[d{\sigma}^{\uparrow} \, + \, d{\sigma}^{\downarrow}]}
\label{weight-ssa}
 \ee
where $d\sigma^\uparrow$ is the differential cross section in $q_T$ or the y variable and 
${\phi}_{q_T} $ and $\phi_S$ are the azimuthal angles of the $J/\psi$ and proton spin respectively. 
The weight factor in the numerator projects out the Sivers asymmetry.
To evaluate asymmetry in $y$ distribution, we substitute 
\bea
d\sigma ^\uparrow \, - \, d\sigma ^\downarrow=\int d\phi_{q_T}\int q_T\>dq_T
\int_{4m^2_c}^{4m^2_D}[dM^{2}]\int[d^2\bfk_{\perp g}]
\Delta^{N}f_{g/p^{\uparrow}}(x_{g},\bfk_{\perp g}) \nonumber \\
\times\>f_{\gamma/e}(x_{\gamma},\bfq_T-\bfk_{\perp g}) \>
\hat\sigma_{0}(M^2) \label{dxsec-y}
\eea
and
\bea
d\sigma ^\uparrow \, + \, d\sigma^\downarrow= 
2\int d\phi_{q_T}\int q_T\>dq_T\int_{4m^2_c}^{4m^2_D}[dM^{2}]\int[d^{2}\bfk_{\perp g}]
f_{g/p}(x_g,\bfk_{\perp g}) \nonumber \\
\times\>f_{\gamma/e}(x_{\gamma},\bfq_T-\bfk_{\perp g})\,
\hat{\sigma}_0(M^2). 
\label{txsec-y}
\eea

Thus at LO, the SSA depends on the Weizsacker-Williams function, 
gluon distribution function and gluon Sivers function. 
Let us recapitulate some of the details of our choices, discussed in more detail in ~\cite{Godbole:2012bx}. For
 $k_{\perp g}$ dependence of the unpolarized PDFs we use a
simple factorized and Gaussian form \cite{Anselmino-PRD72}:
\be
f_{g/p}(x_g,k_{\bot g})=f_{g/p}(x_g)\frac{1}{\pi\langle k^{2}_{\bot g}\rangle} 
e^{-k^{2}_{\bot g}/\langle{k^{2}_{\bot g}\rangle}}.
\label{gauss}
\ee

In addition to this, we need to specify the transverse momentum dependence of the WW functions. 
In principle, it would be interesting to see whether one can derive an expression for this starting 
from the first principle. But at present we use a simple Gaussian form for the  WW function following the 
corresponding one for the unpolarized PDFs. The form we use is given by   
\be
f_{\gamma/e}(x_\gamma,k_{\bot \gamma})=f_{\gamma/e}(x_\gamma)\frac{1}{\pi\langle k^{2}_{\bot \gamma}\rangle} 
e^{-k^{2}_{\bot \gamma}/\langle{k^{2}_{\bot \gamma}\rangle}}.
\label{gauss-g}
\ee 
In fact, in our earlier work, we had used a dipole form as well. However, 
finding our results to be rather insensitive to the assumed form, in this work we have used 
only the Gaussian form and further use comparable values of 
$\langle{k^{2}_{\bot \gamma}\rangle}$ and $\langle{k^{2}_{\bot g}\rangle}$ .

In Ref.\cite{Godbole:2012bx},  we have used the Sivers function  given by\cite{Anselmino2009}  
\be
\Delta^{N}f_{g/p^{\uparrow}}(x_g,{\bfk_{\perp}}_g)=
\Delta^{N}f_{g/p^{\uparrow}}(x_g)\frac{1}{\pi\langle {k}_{\perp g}^2\rangle}\> 
h({\kt}_g)\>e^{-{k}_{\perp g}^2/\langle {k}_{\perp g}^2\rangle}\cos({\phi}_{k_{\perp}})
\label{sivers2}
\ee
where the gluon Sivers function, $\Delta^Nf_{g/p^{\uparrow}}(x_g)$ is defined as 
\be
\Delta^Nf_{g/p^{\uparrow}}(x_g) = 2\,{\mathcal N}_g(x_g)\,f_{g/p}(x_g).\,
\label{dnf}
\ee
Here,  ${\mathcal N}_g(x_g)$ is an $x$-dependent normalization for gluon and \be
h({\kt}_g) = \sqrt{2e}\,\frac{{\kt}_g}{M_{1}}\,e^{-{k}_{\perp g}^2/{M_{1}^2}}\> 
\label{siverskt}
\ee
$M_1$ is the parameter obtained by fitting the recent experimental data corresponding to pion and kaon 
production at HERMES and COMPASS. 

The parametrizations for the quark Sivers function ${\mathcal N}_u(x)$ 
and ${\mathcal N}_d(x)$ have been 
fitted from SIDIS data and are given by \cite{Anselmino:2008sga}
\be 
{\mathcal N}_f(x) = N_f x^{a_f} (1-x)^{b_f} \frac{(a_f + b_f)^{(a_f +
b_f)}}{{a_f}^{a_f} {b_f}^{b_f}} \; .
\label{admfct}
\ee
Here $a_f, b_f, N_f$ for $u$ and $d$ quarks are free parameters 
obtained by fitting the data. However, there is no information available on ${\mathcal N}_g(x)$.
In our analysis, we have used two parametrizations \cite{Boer-PRD69(2004)094025}
\begin{itemize}
\item[(a)] ${\mathcal N}_g(x)=\left( {\mathcal N}_u(x)+
{\mathcal N}_d(x) \right)/2 \;$,
\item[(b)] ${\mathcal N}_g(x)={\mathcal N}_d(x)$.
\end{itemize}
The first choice assumes that the gluon Sivers function is the average of the up and down quark Sivers function 
while the second choice is motivated by the fact that the gluon distribution function is close to the $d$ quark distribution 
function. There can be other more general parametrizations also; however, in this work we use only these two choices.
It may be noted that out of the remaining two choices proposed in Ref.~\cite{Boer-PRD69(2004)094025}, 
one is irrelevant for us and the other choice differs from choice (b)  only in terms of Gaussian width;  
as per the results of Ref.~\cite{Boer-PRD69(2004)094025}  this is not expected to have too much impact, 
and hence we do not include it.

\subsection{Scale evolution of TMDs}

 In our previous analysis, we had chosen the scale of PDFs and 
Sivers function to be ${\hat s}$. The TMD PDFs and Sivers 
function were obtained through DGLAP evolution by considering the 
evolution of the factorized collinear part.  The issue of scale dependence of 
TMDs has attracted a lot of attention in the recent past and a new formalism 
for the $Q^2$ evolution of TMDs has been developed\cite{Collins:2011book, Aybat:2011zv, Bacchetta:2011gx}.  
Based on this TMD evolution, best fits of the  SIDIS Sivers asymmetry 
have been performed and compared with earlier estimates extracted without 
using TMD  evolution\cite{Anselmino:2012aa}.  In this work, we employ the 
strategy used in Ref.~\cite{Anselmino:2012aa} to take into account the 
TMD evolution of PDFs and the  Sivers function and  compare the resulting 
asymmetries in $J/\psi$ production  with our earlier estimates which were 
obtained using DGLAP evolution.  Since we are following the formalism of 
Ref.~\cite{Anselmino:2012aa}, we will be brief in our description of the 
same. The details can be found in Ref.~\cite{Anselmino:2012aa}.

In this formalism, the $Q^2$ evolution of the $\kt$ dependent distribution 
function is given by \cite{Anselmino:2012aa}
\be
\widehat f_{q/p}(x,\kt;Q)=f_{q/p}(x,Q_0)\; R(Q,Q_0) \; 
\frac{ e^{-\kt ^2/w^2}}{\pi\,w^2} \>,\label{unp-gauss-evol} \quad\quad
\ee
where $f_{q/p}(x,Q_0)$ is the usual integrated PDF evaluated at the initial 
scale $Q_0$ and $w^2 \equiv w^2(Q,Q_0)$ is the ``evolving'' 
Gaussian width, defined as
\be
w^2(Q,Q_0)=\langle\kt^2\rangle + 2\,g_2 \ln \frac{Q}{Q_0}\>  \label{wf}.
\ee 
Here, the evolution factor $R(Q,Q_0)$ is the limiting value of a function $R(Q, Q_0, b_T)$ 
that drives the $Q^2$ evolution of TMDs in coordinate space and is given by \cite{Anselmino:2012aa, Aybat:2011zv} 

\be
 R(Q, Q_0, b_T)
\equiv 
\exp \left\{ \ln \frac{Q}{Q_0} \int_{Q_0}^{\mu_b} \frac{\rm d \mu'}{\mu'} \gamma_K(\mu') +
\int_{Q_0}^Q \frac{\rm d \mu}{\mu} 
\gamma_F \left( \mu, \frac{Q^2}{\mu^2} \right)\right\} 
 \label{RQQ0}
\ee
where $b_T$ is the parton impact parameter and 
\be
\mu_b = \frac{C_1}{b_*(b_T)} \>,  \label{mub}
\quad\quad\quad
b_*(b_T) \equiv \frac{b_T}{\sqrt{1 + b_T^2/b_{\rm max}^2}} , 
\ee
with $C_1=2 e^{-\gamma_E}$ and  $\gamma_E=0.577$ ~\cite{Collins:1984kg}.

In the limit $b_T \rightarrow \infty$ , $R(Q, Q_0, b_T )\rightarrow R(Q, Q_0)$
and  $b_*\rightarrow b_{max}$. $\gamma_F$ and $\gamma_K$ are anomalous dimensions that are given, at $O(\alpha_s)$, by
 
\be
\gamma_F(\mu; \frac{Q^2}{\mu^2}) = \alpha_s(\mu) \, \frac{C_F}{\pi}
\left( \frac{3}{2} - \ln \frac{Q^2}{\mu^2} \right)\\
\label{gammaF}
\ee

\be
\gamma_K(\mu) = \alpha_s(\mu) \, \frac{2 \, C_F}{\pi} \> \cdot
\label{gammaK}
\ee

The TMD evolved  Sivers function is given by\cite{Anselmino:2012aa}
\be
\Delta^N \widehat f_{q/p^\uparrow}(x,\kt;Q)=\frac{\kt}{M_1}\,\sqrt{2 e}\,
\frac{{\langle k_S^2\rangle}^2}{\langle\kt^2\rangle}\,\Delta^N f_{q/p^\uparrow}(x,Q_0)\,R(Q,Q_0)\, 
\frac{e^{-\kt^2/w_S^2}}{\pi w_S^4} \>,\label{siv-gauss-evol}
\ee
with
\be
w^2_S(Q,Q_0)=\langle k_S^2\rangle + 2 g_2 \ln \frac{Q}{Q_0}\>  \label{ws}
\ee
where
\be
\frac{1}{\langle k_{S}^2\rangle} =
\frac{1}{M_1^2}+\frac{1}{\langle k_{\perp}^2\rangle}\> .
\label{ksq}
\ee

\be
\langle k_{\perp g}^2\rangle=0.25 ~GeV^2, \quad\quad g_2=0.68, \quad\quad b_{max}=0.5~GeV^{-1}.
\ee
Here, $M_1$ is a best fit parameter \cite{Anselmino:2012aa}, $Q^2=\hat{s}$,and $Q^2_0=1.0~$GeV$^2$~\cite{Collins:1984kg}.

\subsection{Asymmetry in $J/\psi$ production using evolved TMDs}

Using the TMD evolved PDF and Sivers function given in Eqs. (\ref{unp-gauss-evol}) and (\ref{siv-gauss-evol}),
and following the procedure in Ref.\cite{Godbole:2012bx}, we obtain the  expression for the numerator of asymmetry as
\bea
\frac{d^{3}\sigma^\uparrow}{dyd^2\bfq_T}-\frac{d^{3}\sigma^\downarrow}{dyd^2\bfq_T}
=\frac{1}{s}\int_{4m^2_c}^{4m^2_D}dM^2
\Delta^{N} f_{g/p^\uparrow}(x_{g},Q_0)f_{\gamma/e}(x_{\gamma})
\>\sqrt{2e}\>\frac{q_T}{M_1}\nonumber\\
\times R(Q, Q_0)
 \frac{(\langle{k_{S}}^2\rangle)^2\exp[-q_T^2/(w^2_s+
\langle{{k_{\perp\gamma}^2}}\rangle)]
}{\pi[w^2_s+\langle{{k}_{\perp\gamma}^2}\rangle]^2 \langle{{k}_{\perp}^2}\rangle}
\cos({\phi}_{q_T})\>\hat\sigma_{0}^{\gamma g\rightarrow c\bar{c}}(M^2) 
\label{num2}
\eea\cite{Collins:1984kg}
and the expression for the denominator as
\bea
\frac{d^{3}\sigma^\uparrow}{dyd^2\bfq_T}+\frac{d^{3}\sigma^\downarrow}{dyd^2\bfq_T}
=\frac{2}{s}\int_{4m^2_c}^{4m^2_D}dM^2
f_{g/p}(x_{g},Q_0)f_{\gamma/e}(x_{\gamma}) \nonumber\\
\times\>R(Q,Q_0)\> \frac{\exp[-q_T^2/(w^2+\langle{{k}_{\perp\gamma}^2}\rangle)]
}{\pi[w^2+\langle{{k}_{\perp\gamma}^2}\rangle]}
\>\hat\sigma_{0}^{\gamma g\rightarrow c\bar{c}}(M^2) \label{den-m1}.
\eea

We will use these expressions to estimate the asymmetry in the next section. It is known that the Gaussian approximation
is suitable only for low ${\bfk_{\perp}}$  values as the Sivers function develops a tail at high ${\bfk_{\perp }}$ values\cite{Aybat:2011ge}. 
Therefore, while integrating over ${\bfk_{\perp }}$, one needs to set the upper limit of integration below the value 
where the Sivers function starts deviating from the Gaussian form. 
However,  we are justified in integrating over the whole range of ${\bfk_{\perp }}$ of gluon as 
the finite width of the Gaussian dependence of the WW function; in fact, it provides additional damping, making the effective  
range of integration to be less than the value at  which the above-mentioned tail starts.  
We have, in fact,  also  performed  the ${\bfk_{\perp g}}$ integration in Eqs. (\ref{dxsec-y}) and (\ref{txsec-y}) 
numerically over a  finite range allowed by  Gaussian approximation and verified that the result is the same as obtained 
by analytical integration. 

\section{NUMERICAL ESTIMATES  FOR THE ASYMMETRY IN $J/\psi$ PRODUCTION USING TMD EVOLVED SIVERS FUNCTION}

We will now estimate the magnitude of asymmetry using the TMD evolved PDFs and Sivers function and 
compare the results with our previous estimates of asymmetry using DGLAP evolution\cite{Godbole:2012bx}. 
We will also compare our results with asymmetry calculated using 
DGLAP evolution with another set of  parameters extracted from DGLAP fits at $Q_0=1.0$~GeV.
 
For the DGLAP evolution, we have estimated asymmetry using two different sets of parameters. 
The first set,  which we call DGLAP1, consists of the values of the best fit parameters
that we have used in Ref.\cite{Godbole:2012bx}: 
\bea
N_u = 0.40, \ a_u=0.35, \ b_u =2.6 \; , \nonumber \\
N_d = -0.97, \ a_d = 0.44, \ b_d=0.90 \;, \nonumber \\
M_1^2=0.19~GeV^2.
\label{2011-parm}
\eea
These parameters are from new HERMES and COMPASS data \cite{hermes09,compass09} 
fitted at $Q^2 = 2.4~GeV^2$\cite{Anselmino:2011gs}.  

The second set of parameters, which we call DGLAP2, has been extracted from DGLAP fits at $Q_0=1.0~GeV$\cite{Anselmino:2012aa}: 
\bea
N_u = 0.45, \ a_u=1.08, \ b_u =6.9 \; , \nonumber \\
N_d = -1.00, \ a_d = 1.7, \ b_d=6.9 \;, \nonumber \\
M_1^2=0.19~GeV^2.
\label{2012-dglap-parm}
\eea

In both cases, the value of $\langle{k_{\perp g}^2}\rangle$ is chosen to be the  same 
as $\langle{k_{\perp}^2}\rangle $ for quarks obtained in Ref.~\cite{Anselmino:2005nn} by analysis 
of the Cahn effect in unpolarized SIDIS from data collected in different energy and $Q^2$ ranges assuming a 
constant Gaussian width. The value of $\langle{k_{\perp \gamma}^2}\rangle$ 
has been chosen to be 0.25 $GeV^2$ as in our earlier work. 

For TMD evolved Sivers function, we have used the parameter set fitted at $Q_0=1$~GeV given in Ref. \cite{Anselmino:2012aa}. 
\bea
N_u = 0.75, \ N_d = -1.00, \ b=4.0 \; , \nonumber \\
a_u=0.82, \ a_d = 1.36, \ M_1^2=0.34~GeV^2 \;.\label{tmd-parm} 
\eea

We have estimated the asymmetry with both kinds of parametrizations [labeled (a) and (b)].
The estimates are obtained using GRV98LO for unpolarized gluon distribution functions \cite{Gluck:1998xa} and 
the Weizsaker-Williams function for photon distribution \cite{ww}. 

In Figs. 1-6 we have performed a comparative study of Sivers asymmetry $A_N^{\sin({\phi}_{q_T}-\phi_S)}$
for DGLAP evolution and TMD evolution as a function of rapidity $y$ and $q_T$, the transverse momentum of 
$J/\psi$, respectively for JLab ($\sqrt s = 4.7$ GeV), HERMES ($\sqrt s = 7.2 $ GeV), COMPASS ($\sqrt s = 17.33$ GeV) 
and eRHIC ($\sqrt s = 31.6 $ GeV and $\sqrt s = 158.1 $ GeV) energies.
Before we start with a detailed discussion, we make one observation. 
In all the cases the asymmetries are smaller than our earlier estimates. 
However, they still remain sizable, except for the lowest energy at JLab, 
where it is a difficult measurement due to the low event rate.
  
In Figs. 1-5, SSA with DGLAP evolution with parameter sets DGLAP1 and DGLAP2  and TMD evolution is compared 
using parametrization (b). To get the $q_T$ distribution of asymmetry we have integrated over all
possible values of $y$, and to get the $y$ distribution of asymmetry we have integrated over  $q_T$ from $0$ to $1.0$ GeV. 
The masses of the $c$ quark  and the $D$ meson are taken to be $1.275$ and
$1.864$ GeV, respectively \cite{Beringer:1900zz}.  In the TMD evolved PDFs and Sivers function, 
the initial scale $Q^2_0$ has been chosen to be $1.0$ GeV as we are  using Sivers function parameters  
of Ref.~\cite{Anselmino:2012aa} fitted at this scale. 
Since we are using the color evaporation model,  evolved TMDs are evaluated at $Q^2 = \hat{s}$, which  varies  
between $4m^2_c$ and  $4m^2_D$ and is the only relevant scale for $J/\psi$ production in the CEM.  

We note that all the asymmetries are suppressed with respect 
to our earlier predictions using the older model and parameter set DGLAP1\cite{Godbole:2012bx}. To be specific, we see that
for parametrization (b) of the gluon Sivers function, with TMD evolution, the maximum asymmetry in the $y$ distribution is 
reduced from approximately 15$\%$ to 5$\%$ at different rapidity values, 
for energies of JLab, HERMES, COMPASS, eRHIC-1, and eRHIC-2 experiments. 
For the $q_T$ distribution of asymmetry with parametrization (b), the maximum asymmetry is reduced from  
approximately 15$\%$ to 5$\%$ at $q_T$=0.75 GeV for  energies of JLab, HERMES, COMPASS  and eRHIC-1 experiments. 
For higher eRHIC energy, the asymmetry in $q_T$ distribution is reduced from approximately 5$\%$ to 1$\%$.

We have also compared our estimates of asymmetry using TMD evolved PDFs with the asymmetry calculated using DGLAP evolved PDFs 
with the parameter set DGLAP2.  The parameters in these two cases have been extracted from TMD fits at $Q_0=1.0$~GeV  and 
 DGLAP fits at $Q_0=1.0$~GeV respectively, assuming a Gaussian width of 0.5 GeV at this initial scale. In this case also, 
 we find appreciable suppression but the difference between DGLAP and TMD estimates is smaller now. In particular, 
the asymmetry is not changing much for 
the TMD evolution and DGLAP evolution in case of the $y$  distribution of JLab experiments. 
For the $q_T$ distribution of asymmetry with parametrisation (b), maximum asymmetry is reduced from  
approximately 8$\%$ to 5$\%$ for eRHIC-1 experiment. For the $q_T$ distribution of asymmetry in the eRHIC-2 experiment, 
the asymmetry is reduced from approximately 1.5$\%$ to 1$\%$. In general, we find that the estimates using this parameter set are 
closer to the estimates using TMD evolved PDFs as compared to our earlier estimates using a parameter set fitted at 
$Q^2 =2.4 ~GeV^2$. This seems to suggest the importance of the $Q^2$ dependence of the Gaussian width as 
all the earlier DGLAP fits of the Sivers function have been obtained assuming a constant, scale-independent Gaussian width.

In Fig. 6, we have compared the SSA for parametrizations 
(a) and (b) using TMD evolution. As in case of DGLAP evolution, 
we find that the asymmetry is higher for parametrization (b) 
than that for parametrization (a) in the case of TMD evolution as well.

\section{DISCUSSION AND SUMMARY}
In this work, we have given numerical estimates for $J/\psi$ production in low virtuality 
electroproduction at JLAB, COMPASS, HERMES, and eRHIC energies, using the TMD evolved parton 
distribution functions and Sivers function. At leading order, this asymmetry 
is a clean probe  of the gluon Sivers function. We use here the analytic formulation 
for incorporating the scale evolution of the TMDs and calculate the asymmetries in the color  
evaporation model for $J/\psi$ production. We compared these estimates  with  those obtained 
by us previously using a different parametrization wherein the $Q^{2}$ dependence came from 
the DGLAP evolution of the unpolarized gluon densities. The present estimates, while reduced substantially, 
when compared to our earlier ones, are still sizable 
for most experiments.

Noting that the average $Q^{2}$ for our process is $\sim  10 ~GeV^{2}$, the suppression we find is seen  
consistent with the  suppression in going from HERMES to COMPASS energies, obtained in Ref.\cite{Aybat:2011ta} 
in the context of SIDIS. It should also be remembered that in the case of $J/\psi$ production, 
the transverse momentum dependence of the WW function, can also have non trivial effects.
  
In the model under consideration, the amount of asymmetry is affected by the evolution factor $R(Q,Q_0)$ as well as by 
the $Q^2$ dependence of the Gaussian width. In the color evaporation model, $Q^2$ varies between $4m_c^2$ and $4m_D^2$ and
$R(Q,Q_0)$ does not vary much over this range.  Hence, this factor approximately gets canceled between the numerator 
and denominator in the evaluation of asymmetry. Thus, the noticeable suppression comes from the logarithmic dependence on 
$Q^{2}$ of the Gaussian width. This is consistent with the observation in \cite{Anselmino:2012aa}  
that the substantial decrease in asymmetry is mainly due to the $Q^{2}$ dependence of the Gaussian width of TMD PDFs. 
Hence, it is important to understand this $Q^{2}$ dependence of Gaussian width in order to be able to make predictions that
can be compared  with future experiments. 
In the case of $J/\psi$ production, there is yet another convolution with a transverse momentum dependent WW function. 
In principle, one needs to model its scale dependence as well. 
In the end it should be noted that, since the hard scale is the same for all beam energies,
the Sivers function is probed in the same $Q^{2}$ range. This explains the similar amounts of reduction of asymmetries 
seen at all energies.  

Our estimates in the present work are based on the approximate 
form of the Sivers function in Eq.~(\ref{siv-gauss-evol}) which 
was obtained in Ref.~\cite{Anselmino:2012aa} using TMD 
evolution equations of the Refs.~\cite{Collins:2011book,Aybat:2011ge}. This form, called the analytic form in 
Ref.~\cite{Anselmino:2012aa}, is obtained by making an approximation 
$b_T \rightarrow \infty$ in Eq.~(\ref{RQQ0}). An exact form can be obtained from Eq.~(\ref{RQQ0}) by first evaluating the 
TMD evolved PDFs in coordinate space ${\tilde F}(x, b_T;Q)$ and then taking their Fourier transform.  However, the 
approximate form we have used is valid for $b_T \ge 1.0 ~GeV^{-1} $ if the typical $k_\perp $ involved is of the order of 
1 GeV. Using the exact form may lead to slight differences 
from the present estimates, and we plan to address this 
issue in future work.   
 
\section{ACKNOWLEDGEMENTS}

R.M.G. wishes to acknowledge support from the Department of Science and
Technology, India, under Grant No. SR/S2/JCB-64/2007 under the J.C. Bose 
Fellowship scheme. A. Misra and V.S.R. would like to thank the Department of Science and Technology, India,
for financial support under Grant No. SR/S2/HEP-17/2006 and to the Department of Atomic Energy-BRNS, India,
under the grant No. 2010/37P/47/BRNS.
\vspace{1 cm}


\newpage
\begin{figure}
\begin{center}
\includegraphics[width=0.35\linewidth,angle=0]{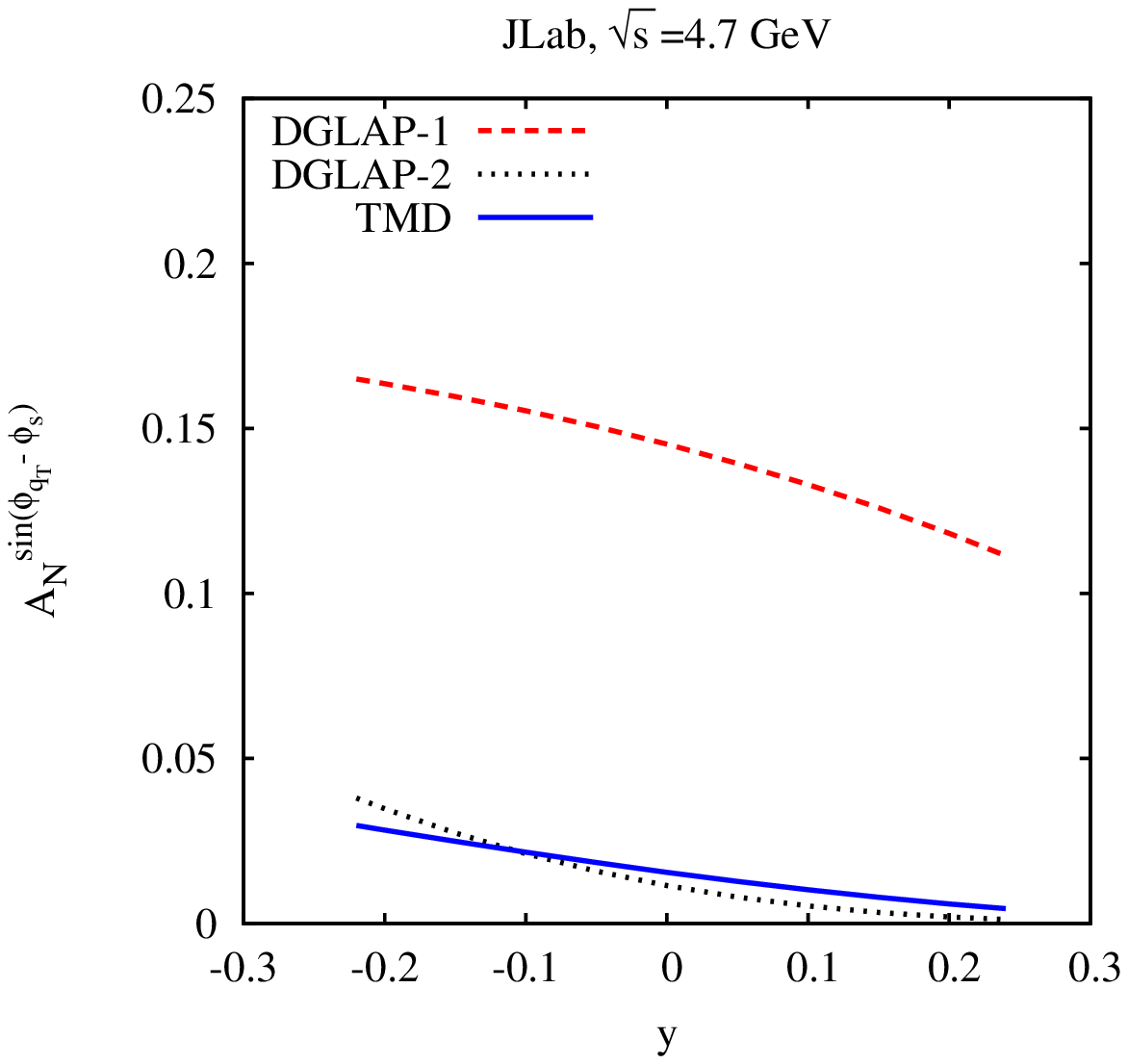}
\includegraphics[width=0.35\linewidth,angle=0]{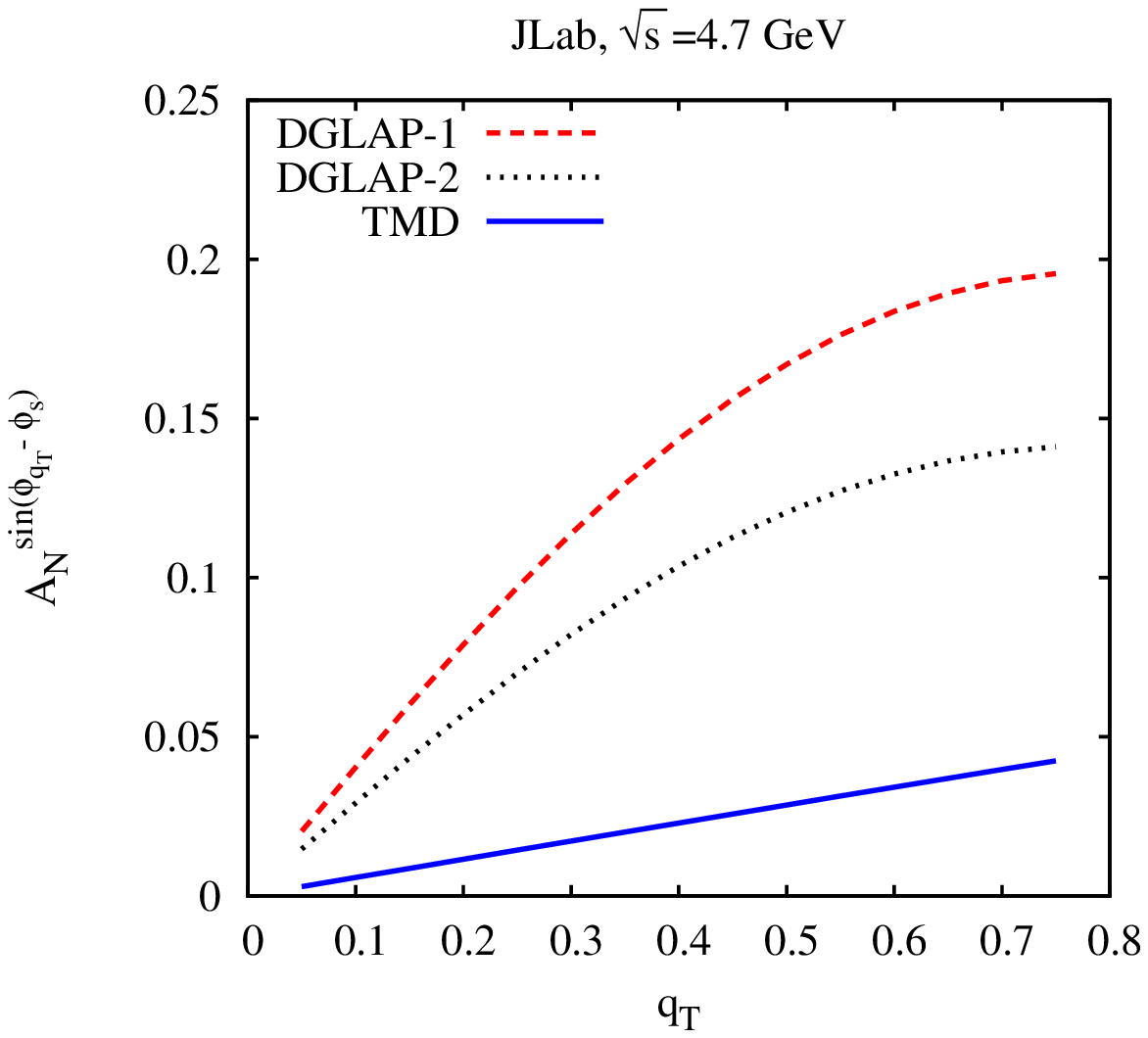}\\
\caption{The Sivers asymmetry $A_N^{\sin({\phi}_{q_T}-\phi_S)}$ for $e+p^\uparrow \to  e+J/\psi +X $
at JLab energy ($\sqrt{s} = 4.7$~GeV) as a function of y (left panel) and $q_T$ (right panel)
for parametrization (b). The solid (blue) line corresponds to results obtained using TMD evolution. 
The dashed (red) and dotted (black) line corresponds to DGLAP evolution with DGLAP fit parameters 
at $Q_0=\sqrt{2.4}$~GeV and $Q_0=1$~GeV respectively . 
The integration ranges are $(0 \leq q_T \leq 1)$~GeV and $(-0.25 \leq y \leq 0.25)$.}
 \vspace{0.5cm}
\includegraphics[width=0.35\linewidth,angle=0]{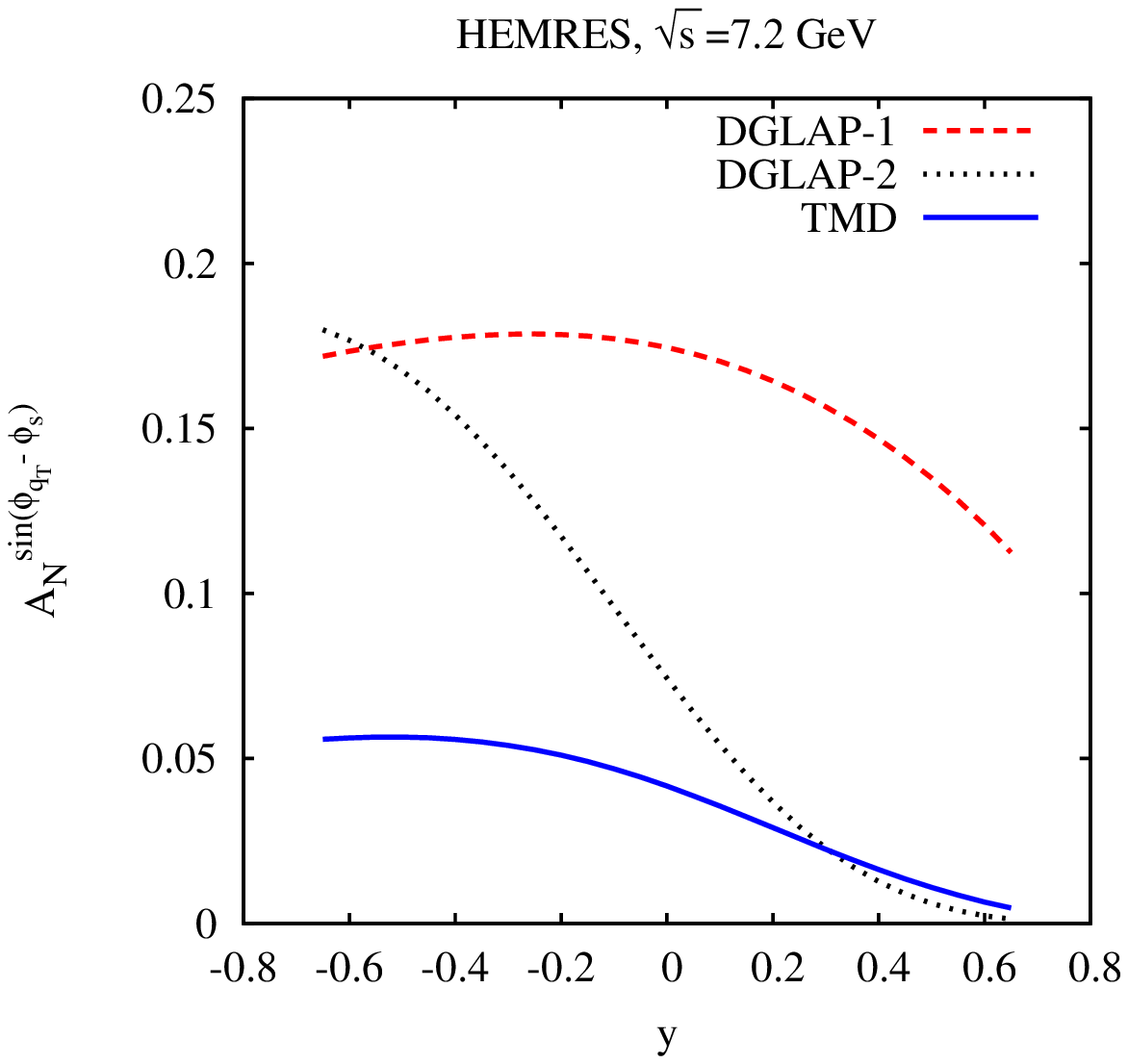}
\includegraphics[width=0.35\linewidth,angle=0]{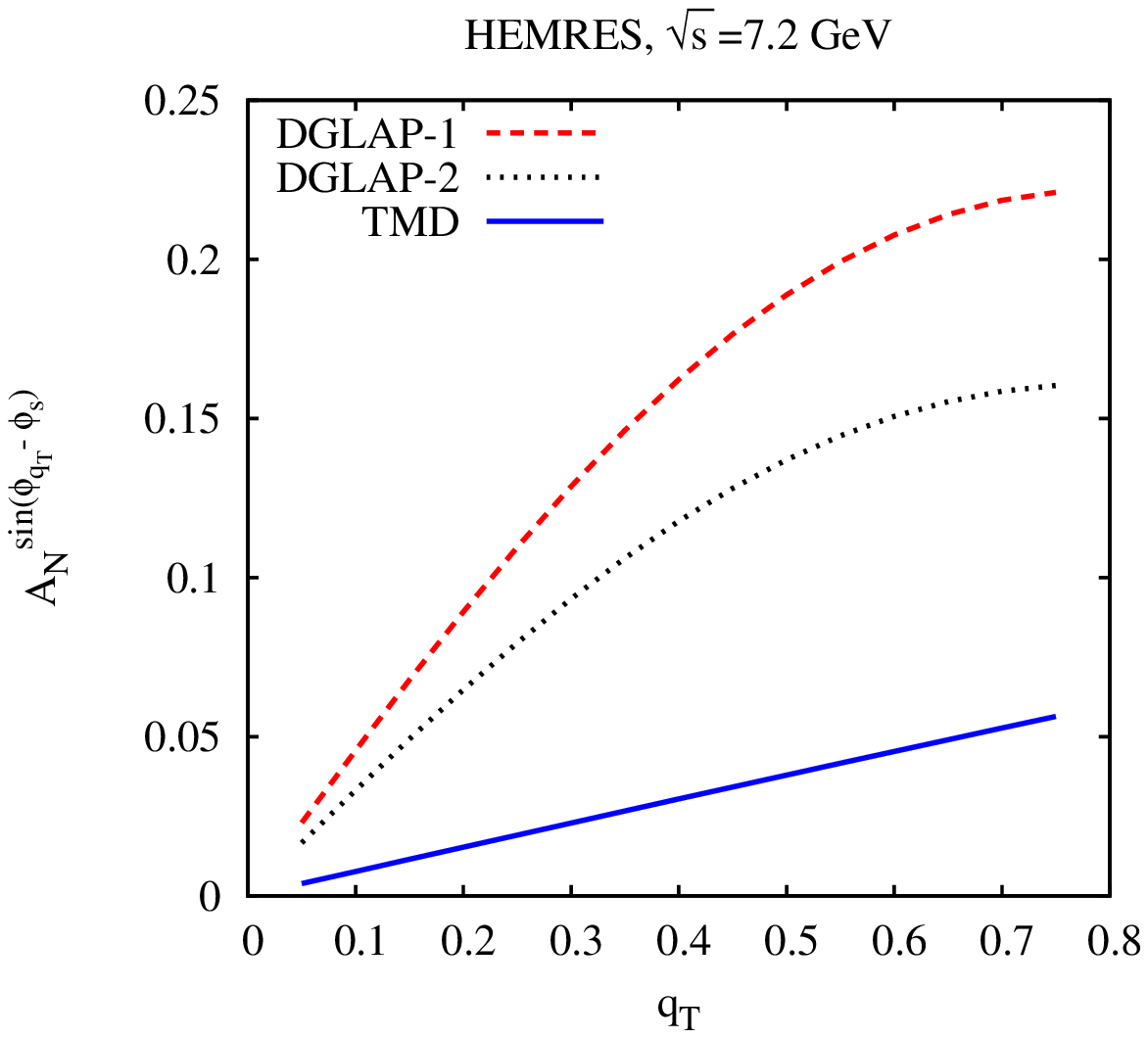}\\
\caption{The Sivers asymmetry $A_N^{\sin({\phi}_{q_T}-\phi_S)}$ for $e+p^\uparrow \to  e+J/\psi +X $
at HEMRES energy ($\sqrt s = 7.2$~GeV) as a function of y (left panel) and $q_T$ (right panel)
for parametrization (b). The solid (blue) line corresponds to results obtained using TMD evolution. 
The dashed (red) and dotted (black) line corresponds to DGLAP evolution with DGLAP fit parameters 
at $Q_0=\sqrt{2.4}$~GeV and $Q_0=1$~GeV respectively . 
The integration ranges are $(0 \leq q_T \leq 1)$~GeV and $(-0.6 \leq y \leq 0.6)$.}
 \vspace{0.5cm}
\includegraphics[width=0.35\linewidth,angle=0]{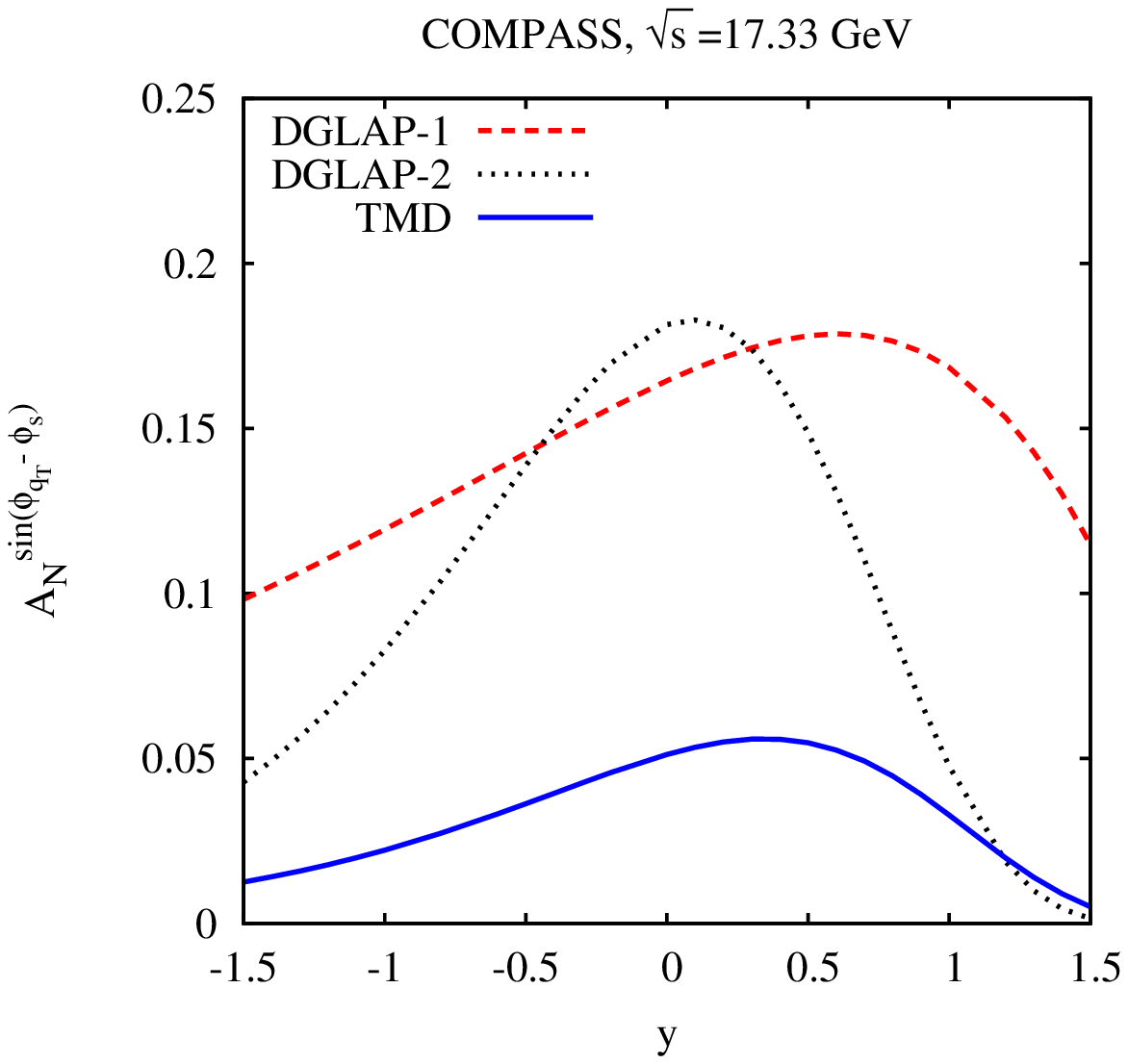}
\includegraphics[width=0.35\linewidth,angle=0]{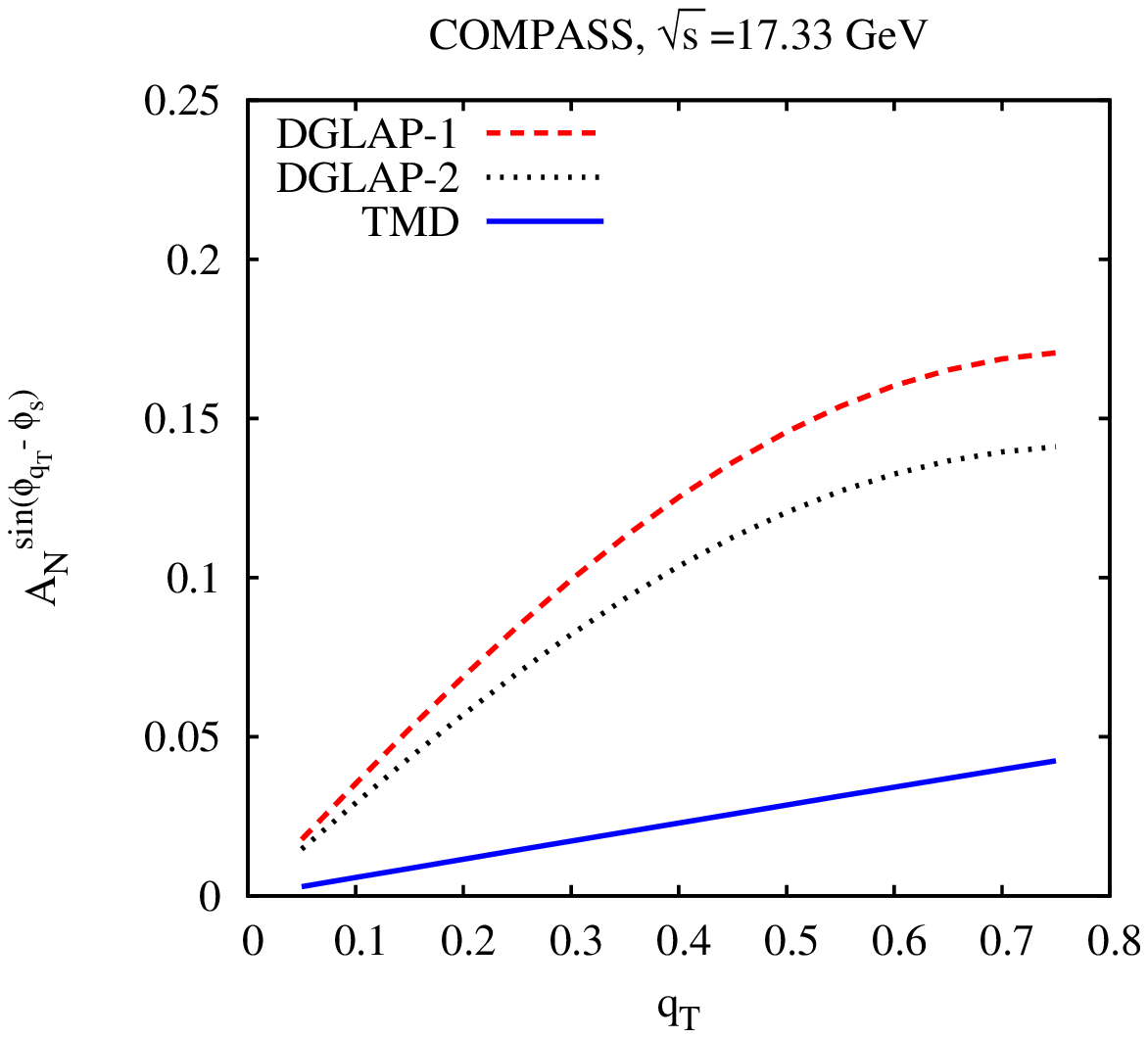}\\
\caption{The Sivers asymmetry $A_N^{\sin({\phi}_{q_T}-\phi_S)}$ for $e+p^\uparrow \to  e+J/\psi +X $
at COMPASS energy ($\sqrt s = 17.33$~GeV) as a function of y (left panel) and $q_T$ (right panel)
for parametrization (b). The solid (blue) line corresponds to results obtained using TMD evolution. 
The dashed (red) and dotted (black) line corresponds to DGLAP evolution with DGLAP fit parameters 
at $Q_0=\sqrt{2.4}$~GeV and $Q_0=1$~GeV respectively . 
The integration ranges are $(0 \leq q_T \leq 1)$~GeV and $(-1.5 \leq y \leq 1.5)$.}
\end{center}
\end{figure}
\begin{figure}
\begin{center}

\includegraphics[width=0.35\linewidth,angle=0]{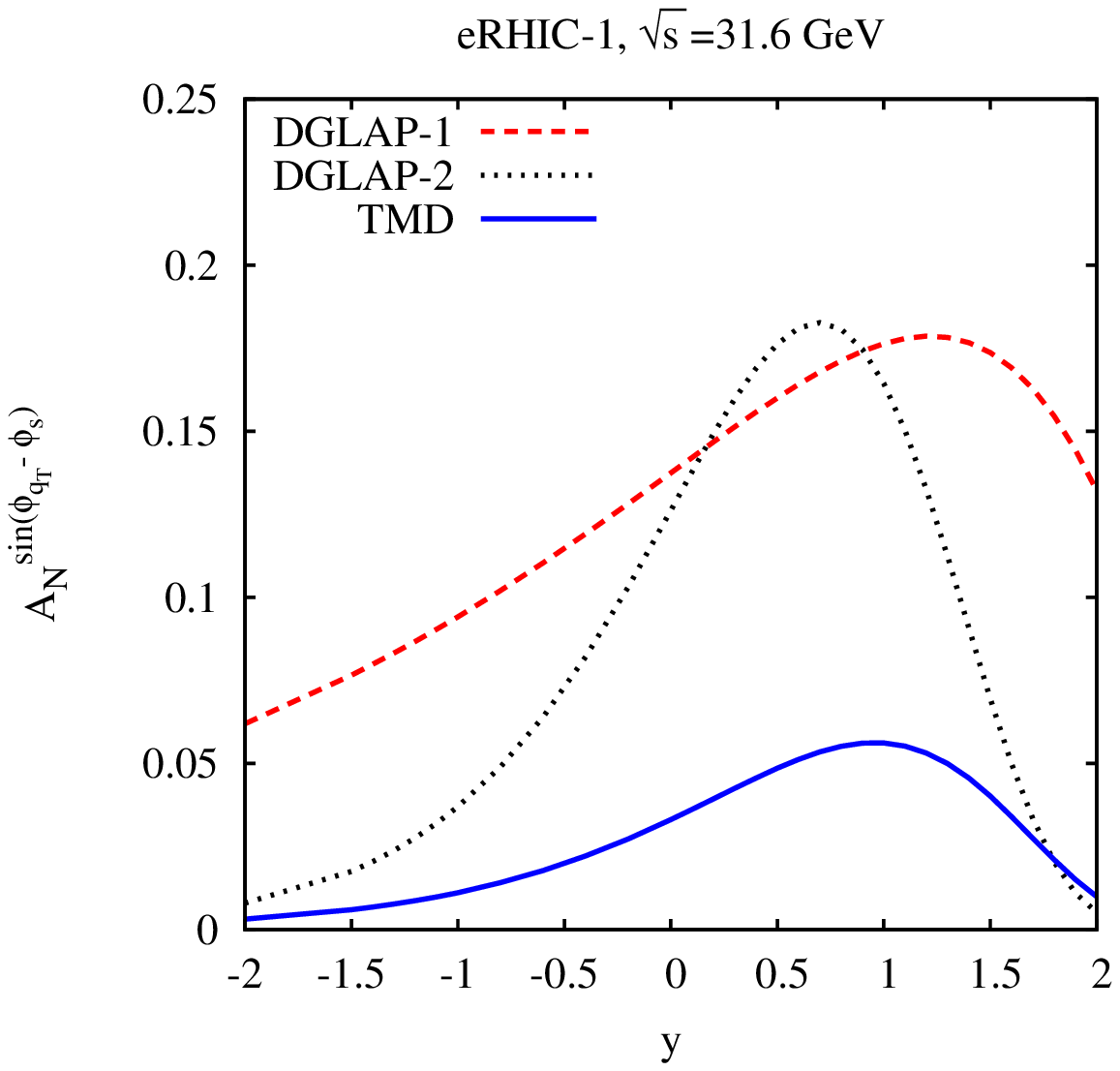}
\includegraphics[width=0.35\linewidth,angle=0]{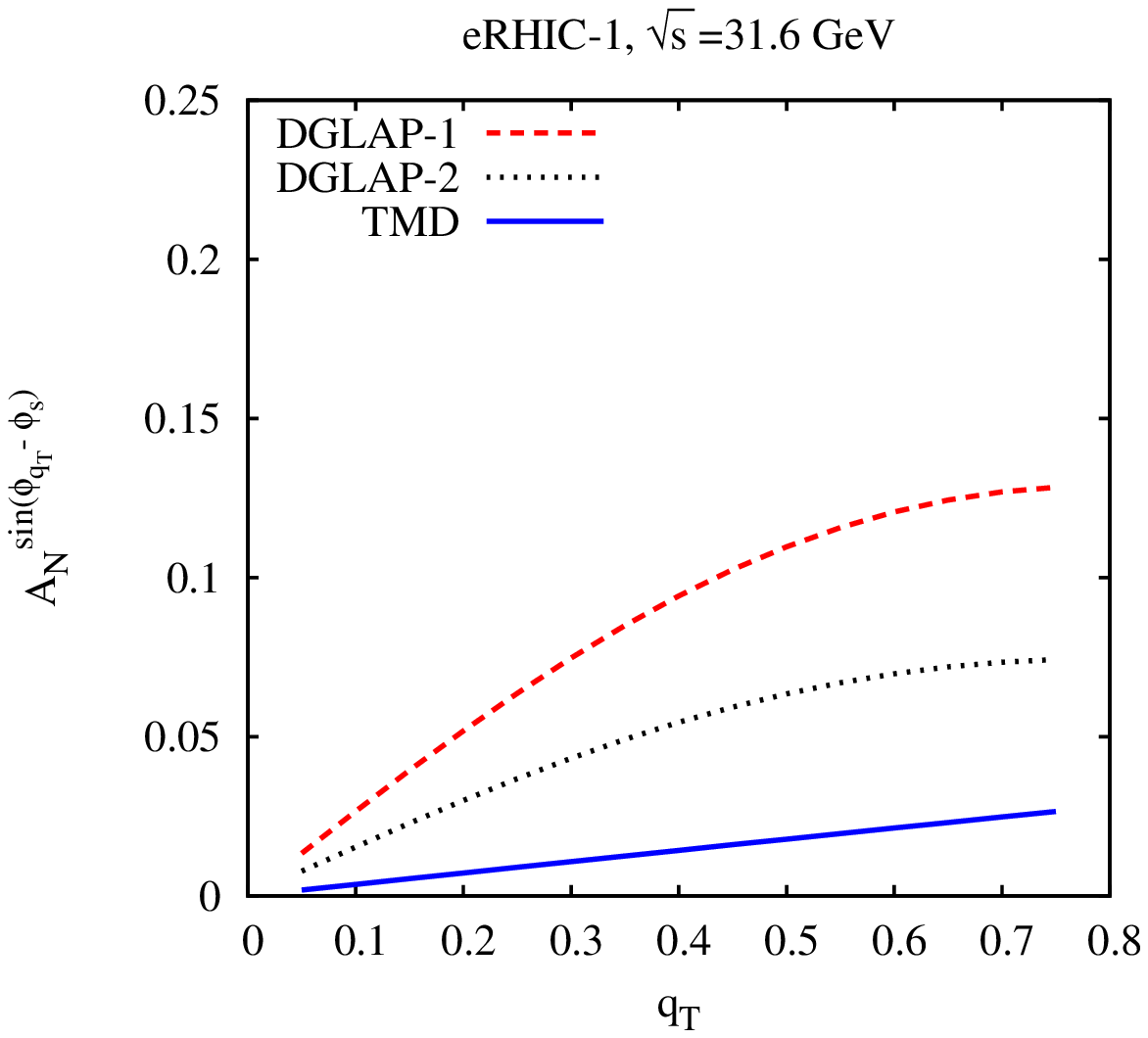}\\
\caption{The Sivers asymmetry $A_N^{\sin({\phi}_{q_T}-\phi_S)}$ for $e+p^\uparrow \to  e+J/\psi +X $
at eRHIC-1 energy ($\sqrt{s}=31.6$~GeV) as a function of y (left panel) and $q_T$ (right panel)
for parametrization (b). The solid (blue) line corresponds to results obtained using TMD evolution. 
The dashed (red) and dotted (black) line corresponds to DGLAP evolution with DGLAP fit parameters 
at $Q_0=\sqrt{2.4}$~GeV and $Q_0=1$~GeV respectively .  
The integration ranges are $(0 \leq q_T \leq 1)$~GeV and $(-2.1 \leq y \leq 2.1)$.}
 \vspace{0.5cm}
\includegraphics[width=0.35\linewidth,angle=0]{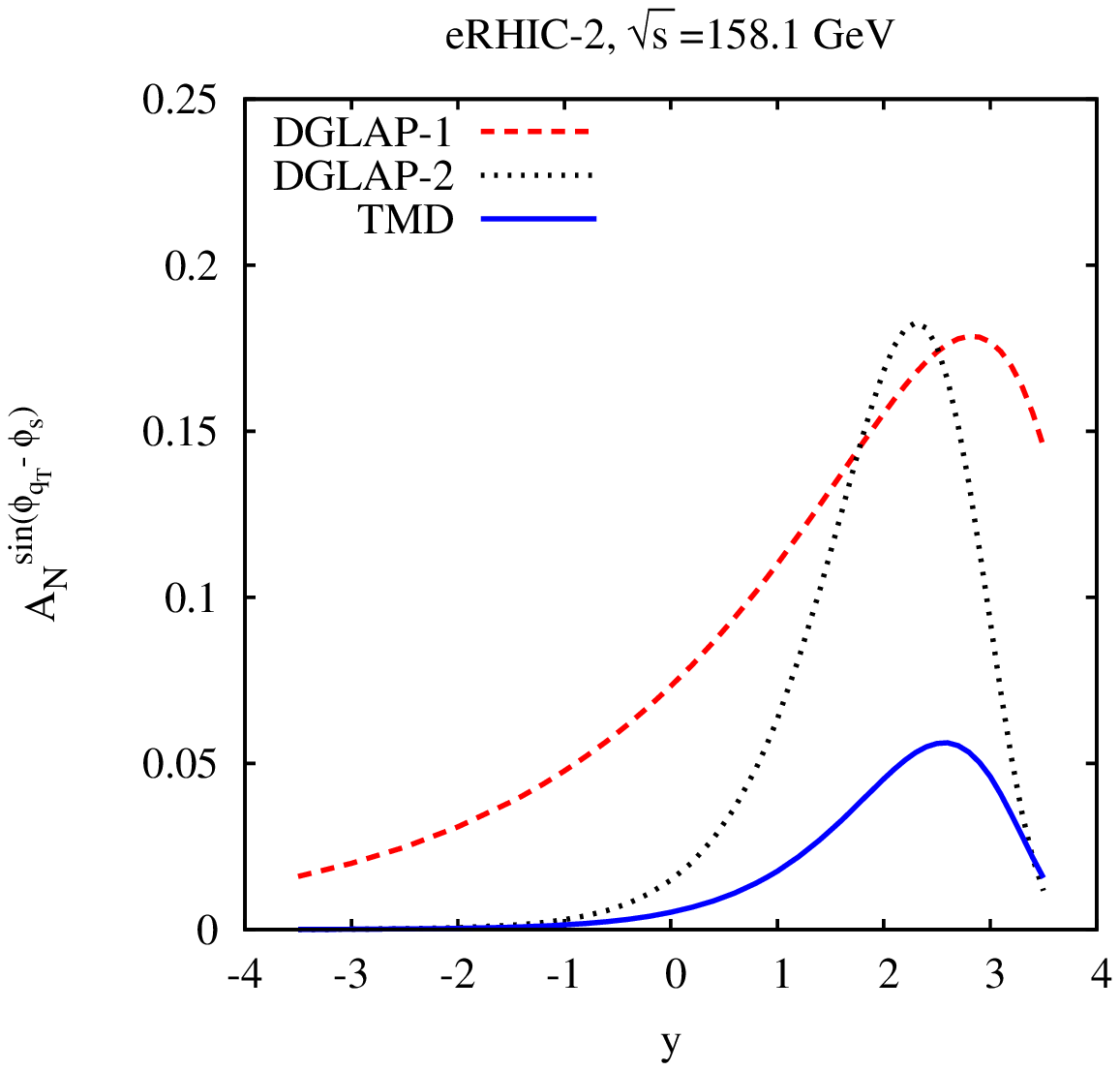}
\includegraphics[width=0.35\linewidth,angle=0]{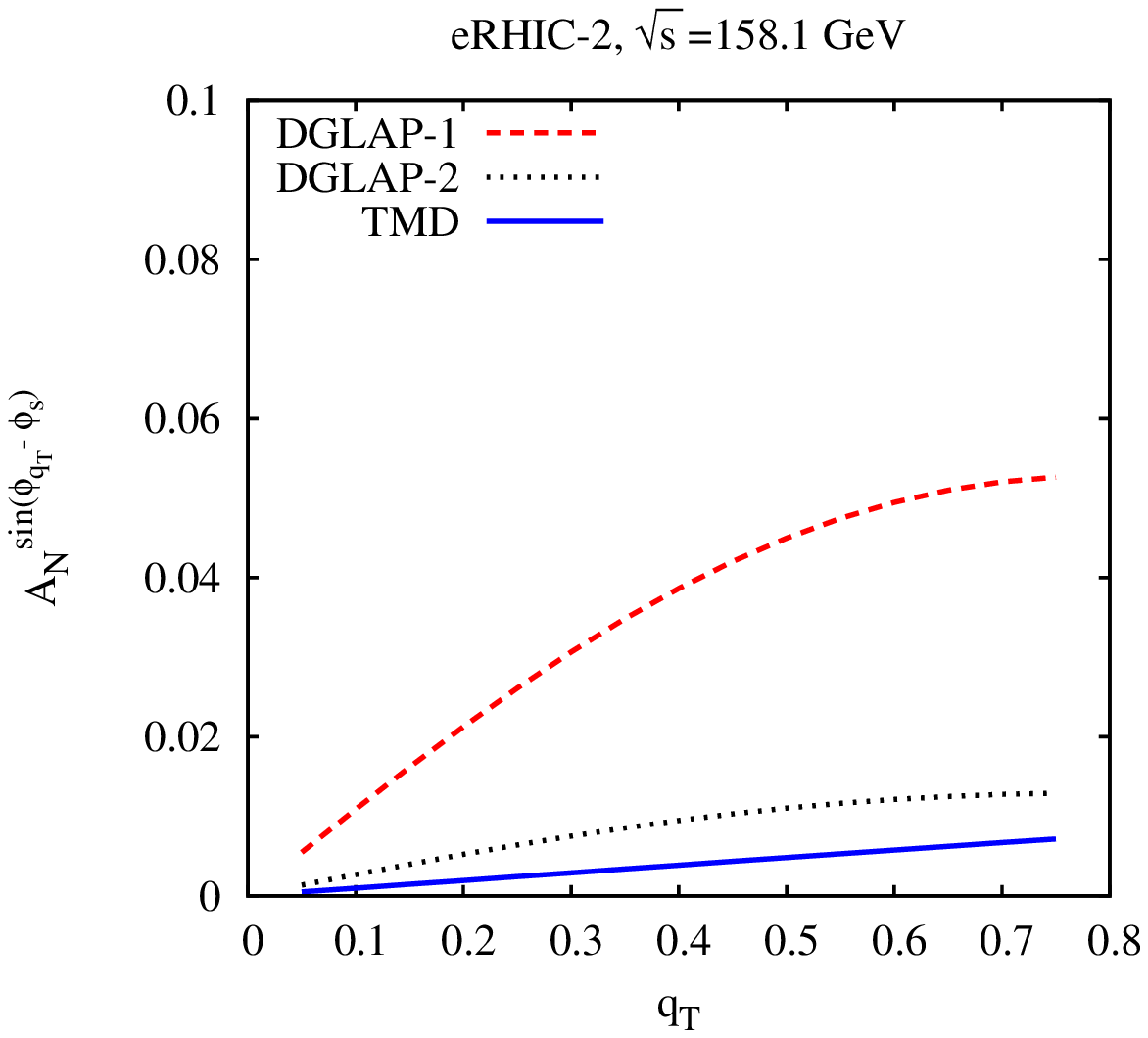}\\
\caption{The Sivers asymmetry $A_N^{\sin({\phi}_{q_T}-\phi_S)}$ for $e+p^\uparrow \to  e+J/\psi +X $
at eRHIC-2 energy ($\sqrt{s}=158.1$~GeV) as a function of y (left panel) and $q_T$ (right panel)
for parametrization (b). The solid (blue) line corresponds to results obtained using TMD evolution. 
The dashed (red) and dotted (black) line corresponds to DGLAP evolution with DGLAP fit parameters 
at $Q_0=\sqrt{2.4}$~GeV and $Q_0=1$~GeV respectively . 
The integration ranges are $(0 \leq q_T \leq 1)$~GeV and $(-3.7 \leq y \leq 3.7)$.}
 \vspace{0.5cm}
\includegraphics[width=0.35\linewidth,angle=0]{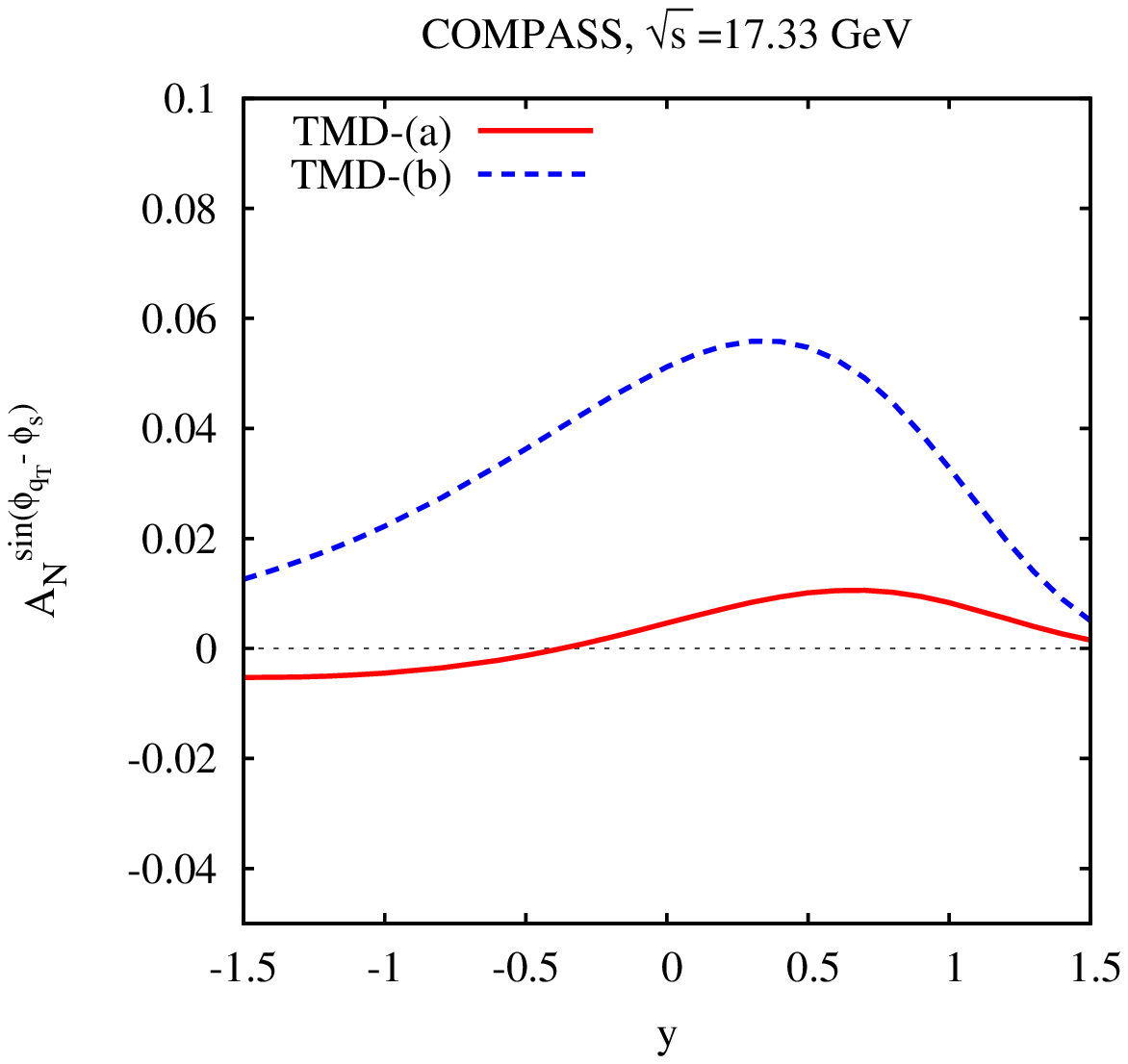}
\includegraphics[width=0.35\linewidth,angle=0]{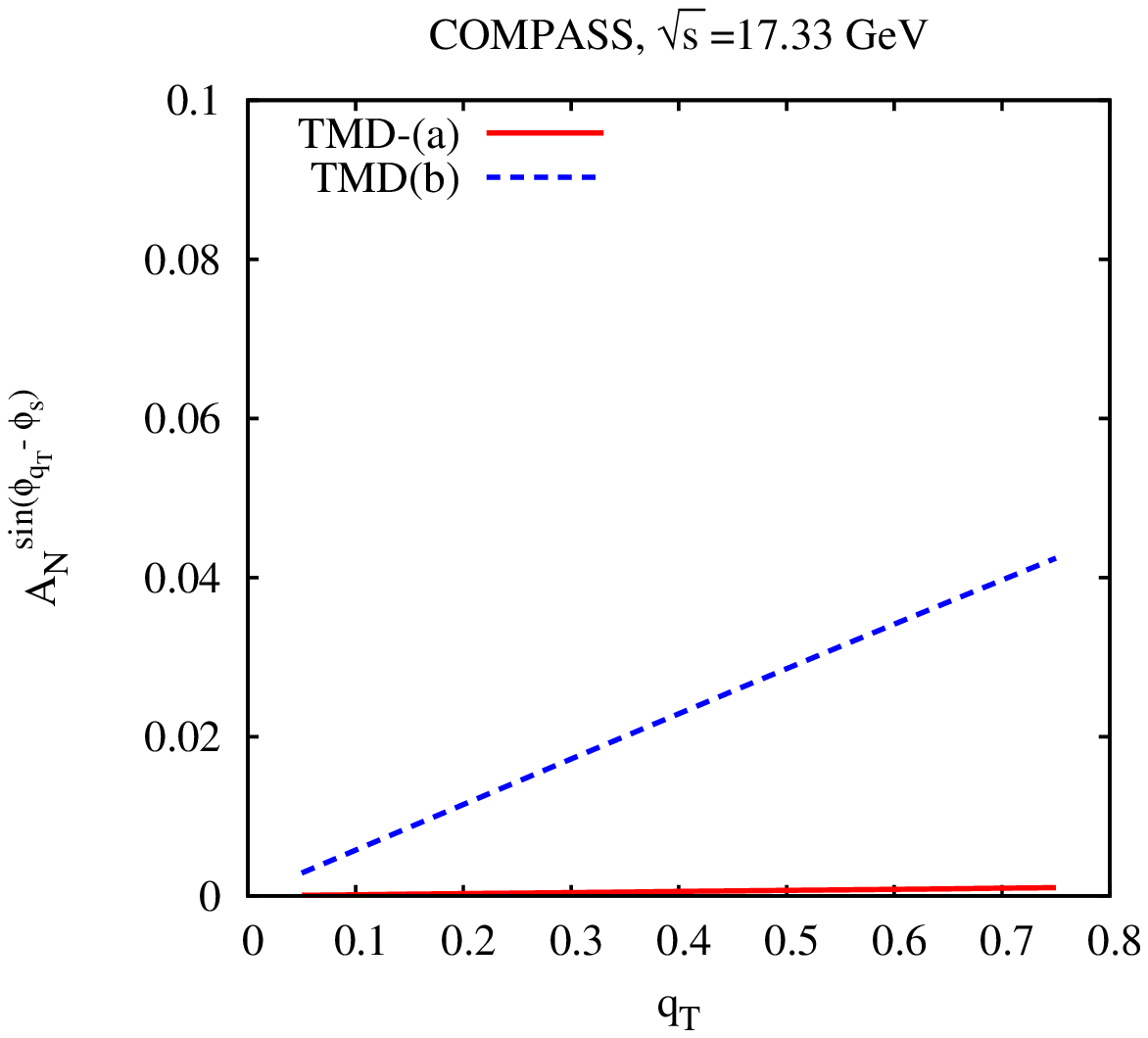}\\
\caption{The Sivers asymmetry $A_N^{\sin({\phi}_{q_T}-\phi_S)}$ for $e+p^\uparrow \to  e+J/\psi +X $
at COMPASS energy ($\sqrt s = 17.33$~GeV) as a function of y (left panel) and $q_T$ (right panel)
using TMD evolution. The solid (red) line corresponds to results obtained using parametrization (a)  
and the dashed (blue) line correspond to parametrization (b).  
The integration ranges are $(0 \leq q_T \leq 1)$~GeV and $(-1.5 \leq y \leq 1.5)$.}
\end{center}
\end{figure} 

\end{document}